# Cryptography for Multi-Located Parties

Subhash Kak

**Abstract:** This note describes some cryptographic issues related to multi-located parties. In general, multi-located parties make it difficult for the eavesdropper to mount the man-in-the-middle attack. Conversely, they make it easier to address problems such as joint encryption and error correction coding. It is easier to implement the three-stage quantum cryptography protocol.

**Introduction**

This note is to describe issues and problems related to cryptography for a network of distributed, multi-located parties. Given that many business and government entities have secure links between their own geographically distributed computers, multi-location is already a fact for many users. Multi-located parties can address the problem of man-in-the-middle attack [1],[2] much more easily without the need for a certification authority. The existence of multiple links between sender and receiver also make it easier to implement joint encryption and error correction coding [3]-[4], secure communication with side information as in watermarking [5]-[8] as well as the implementation of three-stage protocol for quantum cryptography [9],[10].

Normally, the idea of distributed cryptography is for devising sharing of a secret amongst several parties, as in the case of keys of *k* out of *n* officers in a bank that are required to be used simultaneously to open a vault [11],[12]. Here we speak not of this problem but rather of parties that have more than one location. Thus one can thus imagine party A consisting of computers at different locations with secure communication between them that wishes to communicate with another party B that also consists of several computers at different locations with secure communication amongst them. The links between the computers corresponding to A and B are not secure.

The extension of a party into multiple locations may in some cases be a consequence of trustworthy agents. For example, even in classical cryptography A and B are general labels because of the presence of trustworthy proxies. In the case of a battlefield, a side is not necessarily only the general in his tent but it could consist of several individuals who are distributed on the ground.

In the case of multi-located parties, the eavesdropper cannot easily interpose himself between the sender and the receiver since there exist multiple paths for the sender to send the information to the receiver. The presence of these multiple paths makes it possible to develop richer protocols to guarantee security.

If standard cryptography represents a single transformation on data [13]-[17], multi-located parties effectively allow several transformations in sequence by the parties which facilitate authentication. Furthermore, the encrypted data can be split up and sent over different channels and this opens up the various streams for additional processing.

**Cryptography with authenticating agents**

Historically, in the case of business transactions across large distances, A would have an agent at the destination who would help in authenticating the message for B after an initial exchange with B's agent (Figure 1). Perhaps the authentication would consist of B's agent providing evidence of being the true representative of B, and A's agent providing evidence validating A's seal.

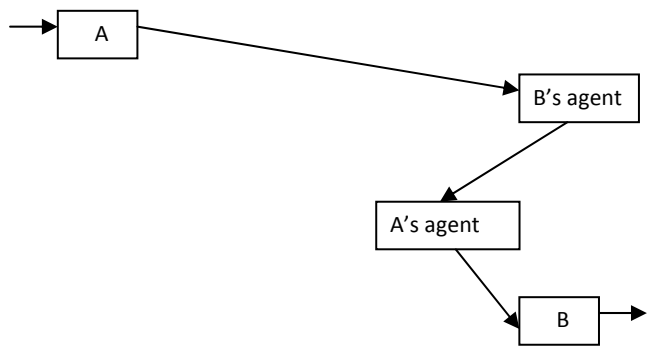

**Figure 1.** Transmission authenticated by agents

If the two parties did not trust each other they could authenticate the transmission by using their own locks (if the contents were in a box that was padlocked from outside) with seals, or by secret transformations on data that they would invert as the message is exchanged between the two parties in the protocol of Figure 2. Proxies [18] could also be used, but that method is not being considered in this paper since it does not include multi-location.

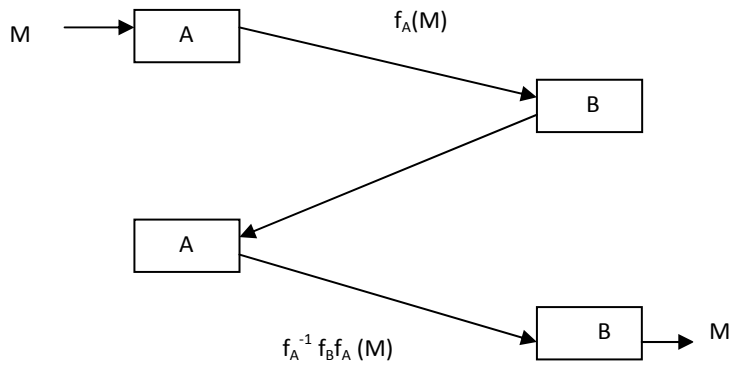

**Figure 2.** Secure three-stage protocol, where $f_B f_A = f_A f_B$

In the 3-stage protocol for secure data transmission over an insecure channel [9],[10] shown in Figure 2, so long as A and B agree to use transformations that commute ($f_B f_A = f_A f_B$), they can use new transformations for each communication (or even each bit) and thus ensure that the communication is unbreakable. The unbreakable nature of this protocol owes to the fact that the entropy of the $f_A$ and $f_B$ processes can be made to equal that of the bit train, if the transformation is viewed to occur on a bit-by-bit basis.

Conversely, one may view the two parties A and B as being distributed, or having the capacity to share information with complete secrecy within each distributed self. We may picture the parties A and B as in Figure 3.

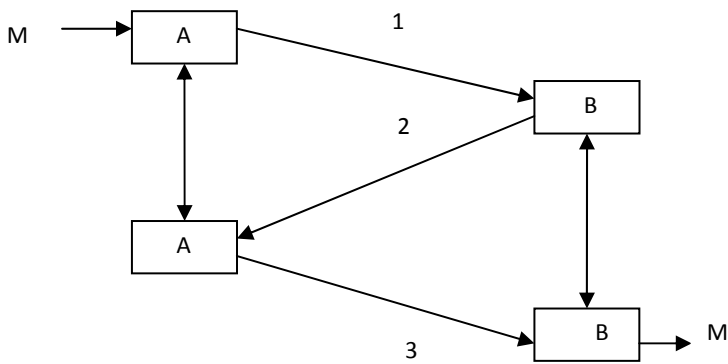

**Figure 3.** Three-stage protocol, with secure connection with agents

Here it is assumed that A and B, while distributed, are located in the same general physical area. If this condition is relaxed and the physical locations of A and B can be interlaced, then one can visualize the scheme of Figure 4 where double arrows are used to show a secure link. The stages 1 through 3 are now each a forward link.

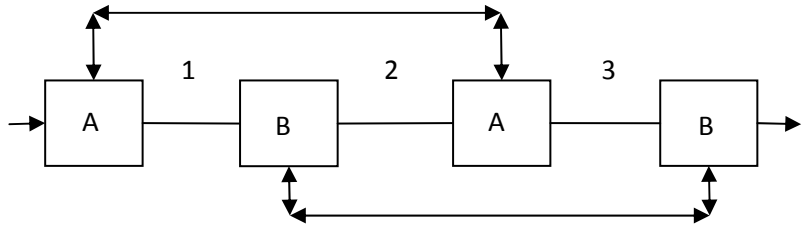

**Figure 4.** Secure three-stage protocol, with distributed parties

Additional units may be visualized in such a chain, and some of these units could be there for error correction.

**The three stage quantum cryptography protocol**

The implementation of the three stage quantum cryptography protocol ordinarily requires, say, sending back and forth of photons with varying polarizations in an arrangement like that of Figure 2. This implementation will become much easier if the photons are sent in a single direction which is the case in the arrangement of Figure 4.

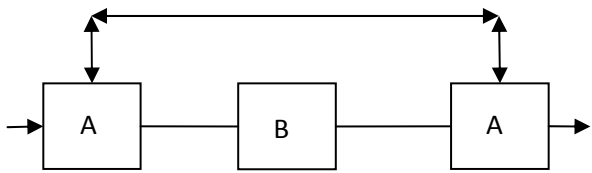

**Figure 5.** A special case of distributed parties

Further simplification may be obtained if the parties are visualized in the special case of Figure 5. This is equivalent to a two-stage cryptography protocol.

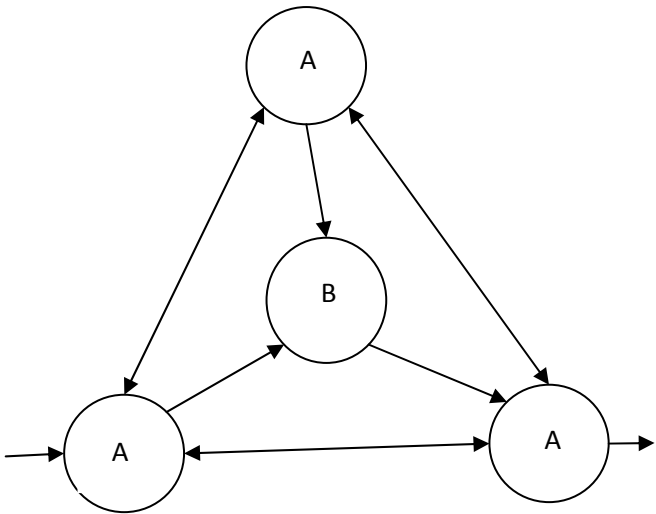

**Figure 6.** Another case of distributed parties

In the example of Figure 6, B receives parts of the encrypted data from two different locations of A and in turn sends a further transformed version to a yet different location of A.

## More general geometries

In a case that is more general than that of Figure 6, both A and B will be multi-located and, therefore, the combinatorial possibilities related to exchange of information between the two will be progressively greater making it corresponding harder for the eavesdropper to break the system.

In the general case the protocol for transmission of information from A to B and the establishment of identity could use specific pathways, some of which are single-directional and others which are bi-directional.

The existence of multiple channels of communication between A and B can also be used to provide redundancy and, therefore, to add error correction capacity to the system. Likewise, in cases where the communication includes side-information [17], different parts of the communication could be sent over different channels. This should make it possible to adapt standard techniques of digital signatures [19],[20] that takes advantage of the multiple channels of communication.

## Discussion

This note describes some cryptographic issues related to multi-located parties. Multi-located parties make it difficult for the eavesdropper to mount the man-in-the-middle attack.

Conversely, they make it easier to address problems such as joint encryption and error correction coding. Such parties make it easier to implement the three-stage quantum cryptography protocol.

The existence of multiple channels between different geometries associated with the two communicating parties makes it possible to devise new classes of security protocols. Space diversity of the transmission process can be exploited by the use of correspondingly richer mathematical structures.